\documentclass[11pt,letterpaper]{article}

\usepackage[T1]{fontenc}
\usepackage[utf8]{inputenc}
\usepackage{mathptmx}
\usepackage{textcomp}
\DeclareUnicodeCharacter{2009}{\,}

\usepackage{geometry}
\usepackage{setspace}

\usepackage{amsmath}
\usepackage{graphicx}
\usepackage{color}
\usepackage{gensymb}
\usepackage{comment}
\usepackage{subcaption}
\usepackage{braket}
\usepackage{float}
\usepackage{authblk}
\usepackage{microtype}

\usepackage[numbers,sort&compress]{natbib}

\title{Tailoring soft cavities for robust molecular strong coupling}

\author[ ]{Siddharaj M. Gadge}
\author[ ]{Adarsh B. Vasista\thanks{Corresponding author: avasista@iiserb.ac.in}}
\affil[ ]{Department of Physics, Indian Institute of Science Education and Research Bhopal, Bhopal, Madhya Pradesh, India}
\date{}

\begin{document}

\maketitle

\begin{abstract}
How should one design efficient chemically open optical cavities for molecular strong coupling? Addressing this question is important for the development of soft-cavity platforms for dynamically tunable light--matter interactions, where direct access to confined electromagnetic modes is essential. Conventional cavity figures of merit such as $Q/\sqrt{V}$ and cooperativity successfully describe spectral confinement and dissipation but do not fully capture the role of linewidth asymmetry between cavity and molecular degrees of freedom. Here, we systematically investigate strong coupling between TDBC dye molecules and whispering gallery modes of polystyrene microspheres by varying the microsphere radius over a broad range. To quantify the robustness of strong coupling, we define the parameter $\chi = \frac{g}{\max(\kappa,\gamma)}$, where $g$ is the coupling strength, while $\kappa$ and $\gamma$ denote the cavity and molecular linewidths, respectively. Although the coupling strength decreases monotonically with increasing cavity size due to mode-volume scaling, we find that $\chi$ exhibits a pronounced maximum near the condition $\kappa \approx \gamma$. This observation suggests that linewidth matching is not merely a criterion for improved spectral visibility, but reflects a dissipation-matching condition that optimizes the robustness of coherent light--matter exchange in soft-cavities. Our results provide an alternative framework for designing morphology-dependent cavities for molecular strong coupling.
\end{abstract}

%%%%%%%%%%%%%%%%%%%%%%%%%%%%%%%%%%%%%%%%%%%%%%%%%%%%%%%%%%%%%%%%%%%%%
%% Main text
%%%%%%%%%%%%%%%%%%%%%%%%%%%%%%%%%%%%%%%%%%%%%%%%%%%%%%%%%%%%%%%%%%%%%

Optical cavities modify the absorption, emission, and energy-transfer processes of molecules through light--matter coupling\cite{10}. In the weak-coupling regime, where the coupling strength, $g$, is smaller than the dissipative rates of the cavity and molecular system, the cavity primarily alters the radiative properties of the emitters, including their emission intensity\cite{1,2,3,4}, polarization\cite{5,22}, and angular distribution\cite{7,8,9}. This regime underlies several enhanced spectroscopic techniques such as surface-enhanced Raman scattering \cite{11,12,13}, surface-enhanced fluorescence\cite{14,15}, etc. In contrast, when the interaction strength overcomes the relevant dissipation channels, the system enters the strong-coupling regime characterized by the formation of hybrid light--matter states\cite{Torma2015}.

Strong molecule--cavity coupling has consequently emerged as a route to influence molecular properties including transport\cite{16,17}, reactivity\cite{18}, and energy-transfer pathways\cite{19}. Among the various cavity architectures explored for molecular strong coupling, microspheres provide a useful platform due to their ability to support spectrally narrow resonances called the whispering gallery modes (WGMs) and chemically open geometry\cite{Vasista2020}. These cavities allow direct access to confined electromagnetic modes while remaining compatible with liquid and microfluidic environments\cite{20}. Importantly, the optical properties of WGMs depend strongly on the microsphere radius, which simultaneously controls the cavity linewidth, mode volume, and field confinement. As a result, the cavity size acts as the principal design parameter governing molecule--cavity interactions in such soft-cavity systems.

The suitability of optical cavities for strong coupling is commonly described using the figure of merit (FOM), $\frac{Q}{\sqrt{V}}$, which captures the balance between spectral confinement through the cavity quality factor $Q$ and spatial confinement through the mode volume $V$ \cite{Chikkaraddy2016}. A more complete description of the coupled system is provided by the cooperativity, $C = \frac{g^2}{\kappa \gamma}$ which incorporates both cavity ($\kappa$) and molecular ($\gamma$) dissipation linewidths \cite{21}. However, cooperativity and FOM does not fully describe situations where the cavity and molecular linewidths differ substantially, as frequently encountered in molecular systems. Thus we have an information gap in designing optical cavities for efficient and robust molecular strong coupling.

In this letter we numerically study J -aggregated 5,5',6,6'-tetrachloro-1,1'-diethyl-3,3'-di(4-sulfobutyl)-benzimidazolocarbocyanine (TDBC) coated microspherical soft-cavities with an intention of achieving robust and efficient strong coupling. To address this question, we introduce the parameter

\begin{equation}\chi = \frac{g}{\max(\kappa, \gamma)},\end{equation}

which compares the coherent interaction strength directly to the dominant dissipation channel. We find that, in the case of soft-cavities, $\chi$ reaches a maximum near the condition $\kappa \approx \gamma$, indicating that balanced photonic and molecular dissipation optimizes the robustness of strong coupling. Thus this criterion naturally defines an optimal cavity size for observing coherent light--matter hybridization. Experimentally, it is well known that matching the cavity and molecular linewidths improves the visibility of mode splitting and polaritonic features. We show that this linewidth matching condition reflects a deeper physical principle beyond spectral observability.

\begin{figure}[htbp]
  \centering
  \includegraphics[width=0.7\textwidth]{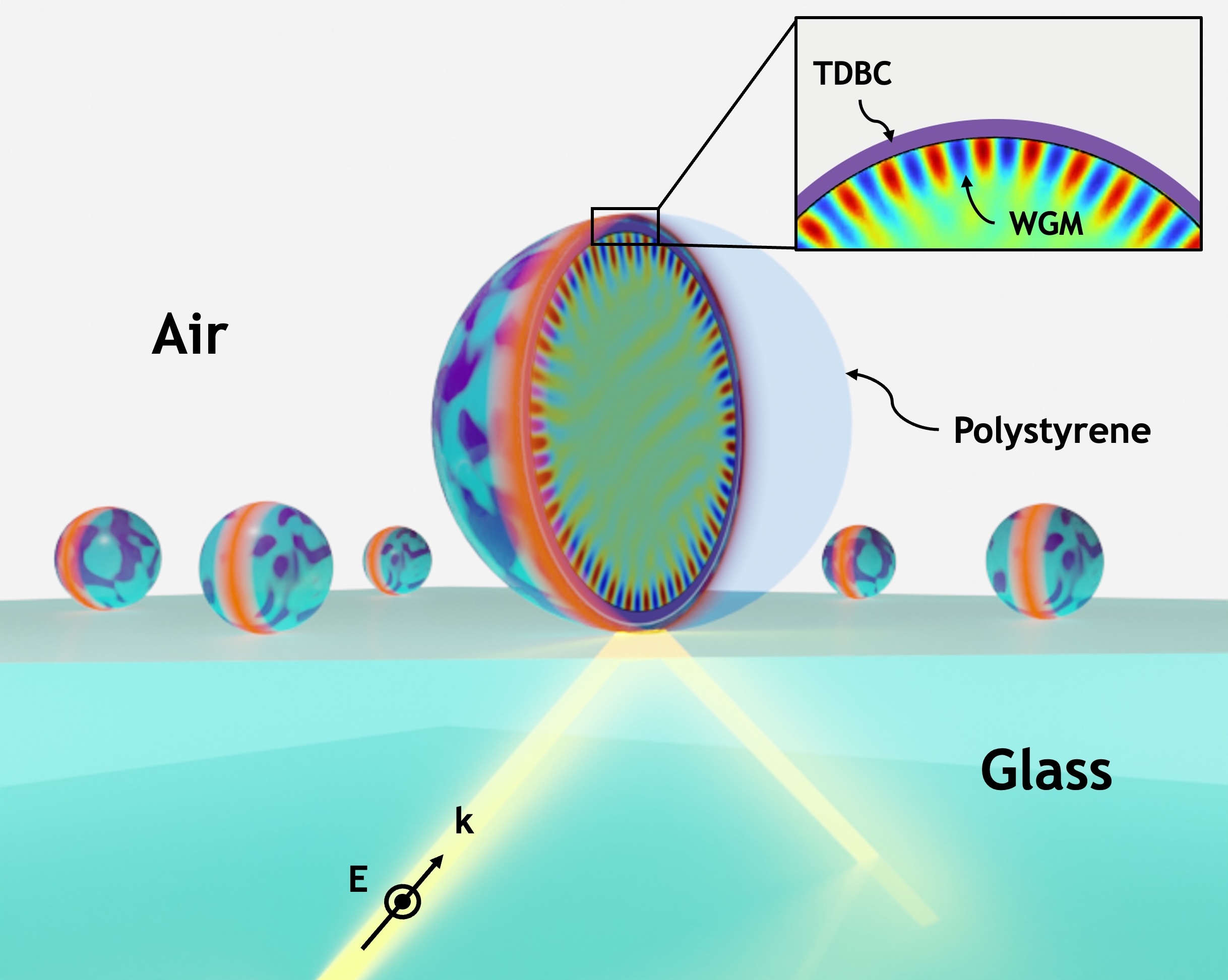}
  \caption{Schematic illustration of the molecule-cavity system under study. The TDBC coated polystyrene microsphere was placed on a glass substrate and probed using evanescent excitation with broadband light.}
  \label{schematic}
\end{figure}

To simulate the hybridized microsphere-molecule system, we developed a two-dimensional (2D) finite element method (FEM) model using COMSOL Multiphysics. Figure \ref{schematic} illustrates the computational domain, which replicates the geometry of a polystyrene microsphere positioned on a glass substrate. We excite the WGMs of the microsphere through the evanescent field generated using the Kretschmann geometry (see section S1, SI for further details on the simulation geometry). As WGMs are morphology dependent resonances, their mode structure depends entirely on the size (diameter) of the microsphere, refractive index contrast between the microsphere and the environment, and the polarization of the exciting light\cite{Vasista2020,Symmetry2022}. The resonances supported by microspheres can be classified as transverse electric (TE) and transverse magnetic (TM) modes characterized by the polarization structure of the electric field. For this computational study, we explicitly chose to excite TE modes of the sphere \cite{Vasista2020}.

WGMs inherently exhibit intense electric-field localization at the physical periphery of the microcsphere (see section S1, SI), ensuring that any layer(s) of molecular J-aggregates coated on the microsphere boundary experience an enhanced electric field (see inset, figure \ref{schematic}). To simulate strong coupling and the resultant light-matter hybridization, we implemented a uniform, concentric thin shell of thickness 8~nm at the microsphere exterior, corresponding to four layers of TDBC, in COMSOL \cite{vasista_2020_SI_ref1}. The complex permittivity of this 8~nm active layer was modeled using a Lorentz oscillator to represent the dispersive macroscopic absorption of the TDBC dye (see section S2, SI). This oscillator was parameterized with an excitonic resonance peak at $\lambda_{res} = 582$~nm and an intrinsic resonance linewidth of $\gamma_{exc} = 43$~meV \cite{vasista_2020_SI_ref1}.

\begin{figure}[htbp]
  \centering
  \includegraphics[width=0.9\textwidth]{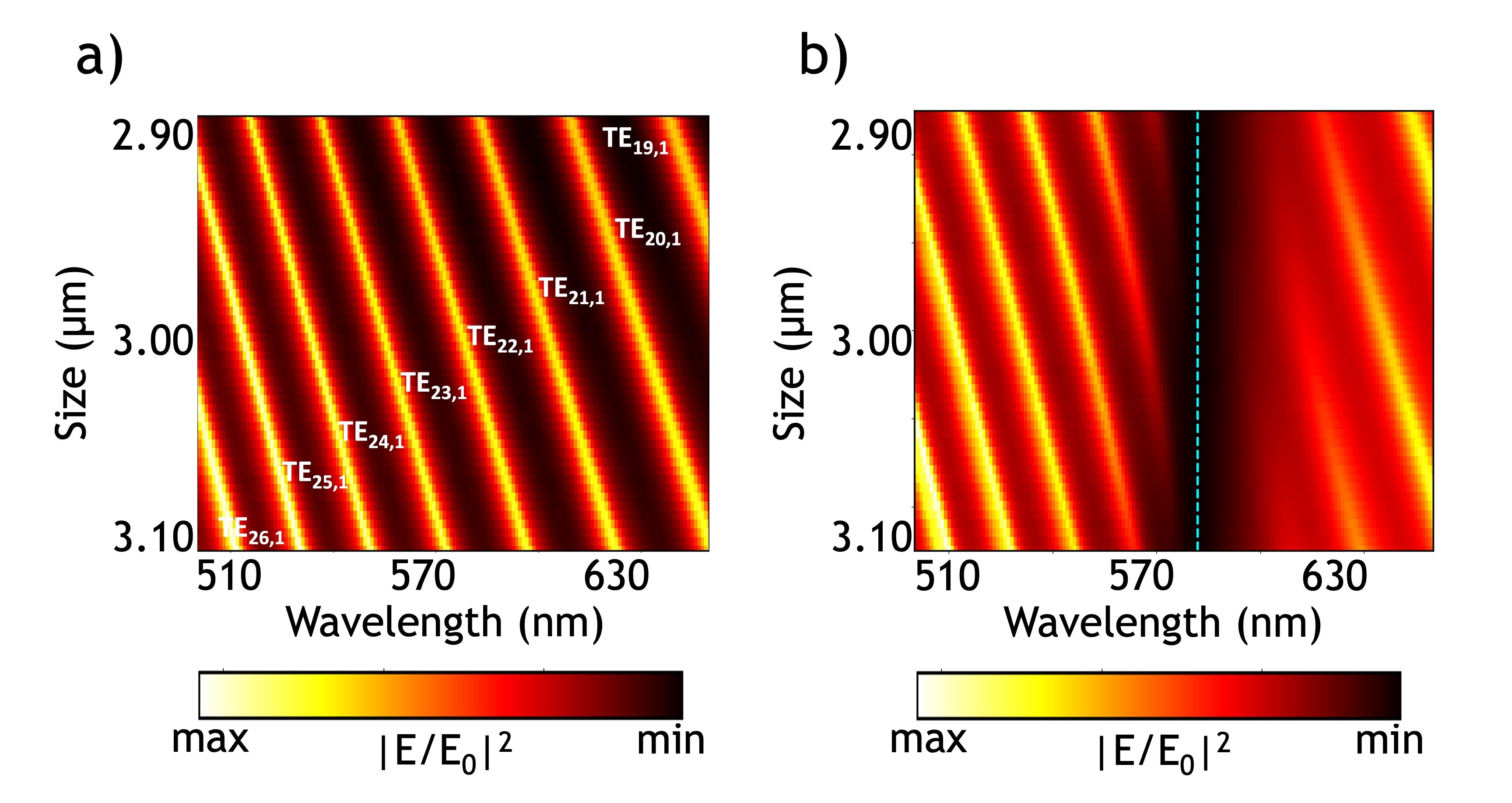}
  \caption{(a) Numerically calculated dispersion of WGMs for a bare micrsophere of nominal size 3 $\mu m$ showing TE modes. (b) Numerically calculated disperion of WGMs for a micosphere coated with 4 layers of TDBC dye molecules showing the splitting and anitcrossing near the absoprtion maxmimum of TDBC (dashed cyan line), a clear signature of molecular strong coupling.}
  \label{fig:dispersion}
\end{figure}

Unlike planar cavities where the cavity resonances can be tuned via the incident wavevector, soft-cavities are morphology dependent resonators\cite{Vollmer2012}. Their resonant frequencies are locked by the physical boundary condition. Therefore, to sweep the cavity energy ($E_{cav}$) across the stationary excitonic resonance ($E_{exc}$) and map the dispersion space to understand and quantify strong coupling, we parametrically swept the diameter of the microsphere in the simulated geometry, keeping the TDBC layer thickness constant. Figure \ref{fig:dispersion} presents the computationally extracted optical response for a bare (uncoated) and dye coated microsphere of nominal diameter 3.0~$\mu$m, swept from 2.90 to 3.10~$\mu$m in fine discrete steps. The color maps represent the 2D surface integration of the internal electric field magnitude ($|E/E_0|^2$), which serves as a robust proxy for the stored electromagnetic energy and resonant scattering cross-section\cite{Vasista2020}.

Figure \ref{fig:dispersion} (a) shows the bare (uncoated) cavity modes. The fundamental TE modes, indexed from $\text{TE}_{26,1}$ through $\text{TE}_{19,1}$ (see section S1, SI for details on indexing of the WGMs), exhibit a linear, dispersive spectral shift as a direct function of the expanding microsphere diameter. Figure \ref{fig:dispersion} (b) displays the computationally extracted dispersion for the identical microcavity geometry coated with the 8~nm TDBC molecular shell. When this active layer is introduced, its near-field interaction with the confined cavity field drastically alters the optical response. Visually, we observe a reduction in the integrated internal electric field magnitude appearing as a dark, attenuated vertical band centered around the TDBC absorption maximum (denoted by the dashed cyan line). This pronounced reduction in mode intensity is a direct consequence of the finite spectral width and the intrinsic dissipation rate ($\gamma_{exc} = 43$~meV) of the molecular absorption band, which introduces significant scattering and absorption losses into the cavity.

As the bare cavity modes dispersively sweep across this fixed molecular resonance, we observe a distinct splitting of the WGMs. The original, continuous uncoupled cavity modes are interrupted, manifesting as a clear avoided crossing near 582~nm. At this stage, the optical response provides direct visual evidence that the dye layer is strongly hybridizing with the cavity field, bifurcating the uncoupled states into distinct upper and lower polaritonic branches.

\begin{figure}[htbp]
  \centering
  \includegraphics[width=\textwidth]{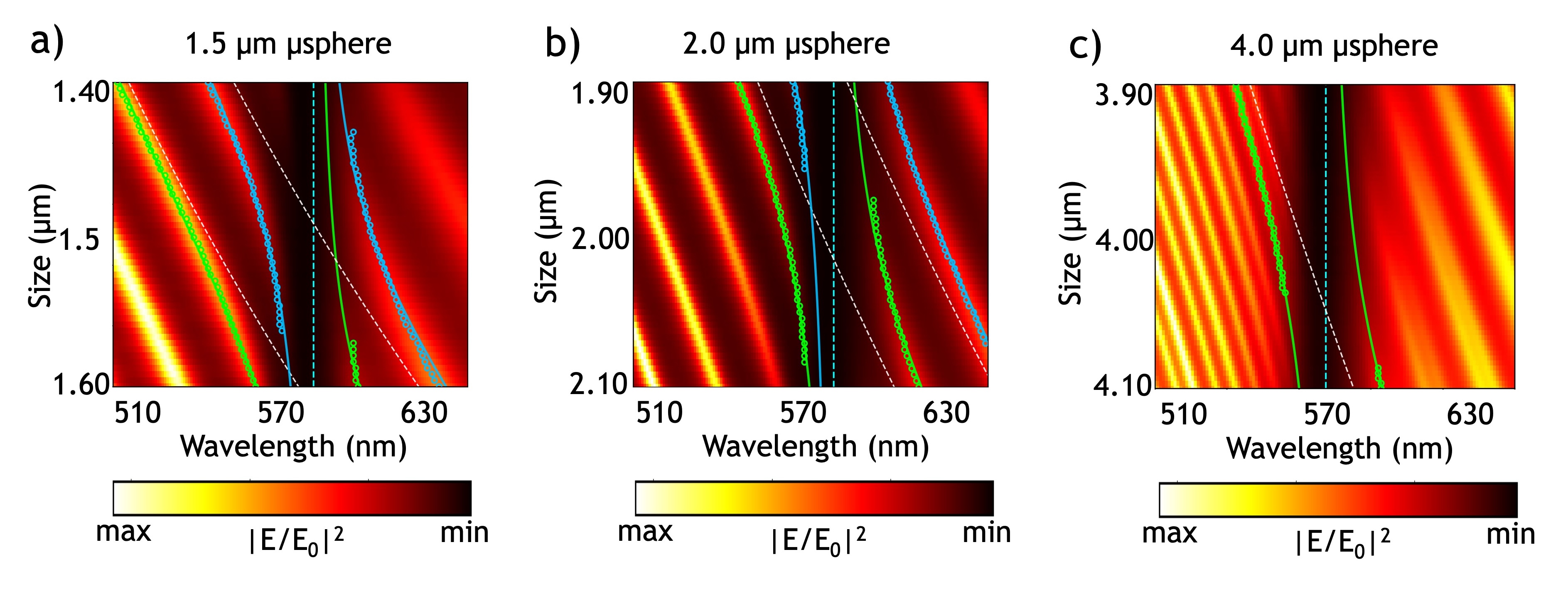}
  \caption{Fitted dispersion plots of 4 layer TDBC coated microspheres of nominal size (a) 1.5 $\mu m$, (b) 2 $\mu m$, and (c) 4 $\mu m$ respectively. The dashed white and cyan lines represent the bare cavity modes and the molecular resonance respectively. The blue and green dotted circles are the fitted polariton maxima from the simulated dispersion plots and the corresponding solid lines are a fit using the coupled oscillator model.}
  \label{fig:com_fits}
\end{figure}

The polariton energies, that is, the eigenenergies of the combined hybrid system, can be approximated from a semi-classical coupled oscillator model (COM)\cite{Vahala2003}. Because whispering gallery mode (WGM) microcavities possess a narrow free spectral range, the molecular resonance often interacts with multiple adjacent cavity modes simultaneously resulting in multi-modal strong coupling\cite{Menghrajani2024}. The Hamiltonian of the total system can be written as:

\begin{equation}
H = \bigoplus_{j} \begin{pmatrix} E_{c,j} - i\frac{\kappa_j}{2} & g_j \\ g_j & E_{exc} - i\frac{\gamma_{exc}}{2} \end{pmatrix}
\end{equation}

where $E_{c,j}$ represents the energy of the $j$th WGM of the resonator and $\kappa_j$ represents its bare cavity linewidth. The parameter $g_j$ is the strength of coupling between the $j$th WGM and the molecular layer, $E_{exc}$ is the molecular absorption energy (centered at $\sim 582$~nm, or $2.13$~eV), and $\gamma_{exc}$ is the linewidth of the molecular absorption.

The eigenvalues of this coupled oscillator matrix give the energies of the upper and lower polariton branches, while the eigenvectors provide an estimate of the mixing fractions, or Hopfield coefficients ($|\alpha|^2, |\beta|^2$), which measure the extent of mixing of the cavity and molecular components of the polaritonic states (see section S3, SI). By fitting this model to the numerically extracted local field maxima from the simulated dispersion maps, we deterministically extract the values of $g_j$. The macroscopic Rabi splitting at zero detuning ($\delta = 0$) is given by the expression:

\begin{equation}
\hbar\Omega_{R,j} = \sqrt{4g_j^2 - (\gamma_{exc} - \kappa_j)^2}
\end{equation}

To systematically investigate how these parameters evolve as a function of the cavity geometry, we performed a comprehensive sweep of the microsphere diameter from $1.5~\mu$m to $4.0~\mu$m in steps of $0.5~\mu$m. Figure \ref{fig:com_fits} presents representative numerically extracted dispersion maxima (open circles) overlaid with the optimized COM eigenvalue fits (solid lines). For the 1.5~$\mu$m microsphere (Figure \ref{fig:com_fits}a) the modes ($\text{TE}_{11,1}$ and $\text{TE}_{10,1}$) exhibit bare cavity linewidths of $\kappa = 64.71$~meV and 78.89~meV, and the fitted coupling strengths were found to be $g = 79.14$~meV and 81.38~meV, respectively, driving a massive Rabi splitting ($\hbar\Omega_R$) of 156.7  and 158.7~meV.

As the diameter increases to 2.0~$\mu$m (Figure \ref{fig:com_fits}b), the interacting $\text{TE}_{14,1}$ and $\text{TE}_{13,1}$ modes show reduced cavity linewidths of $\kappa = 40.62$~meV and 47.74~meV. The expanded mode volume yields slightly lower coupling strengths of $g = 69.07$~meV and 69.47~meV respectively. By 4~$\mu$m (Figure \ref{fig:com_fits}c), interacting with the $\text{TE}_{30,1}$ mode, the cavity linewidths further reduce to $\kappa = 8.7$~meV with coupling strengths evaluated at $g = 23.25$~meV. It is interesting to note that for the microsphere of diameter 4 $\mu$m WGMs of higher radial order (eg., $\text{TE}_{26,2}$) can also be excited within the visible range. For the sake of uniformity we compare the coupling strengths only for radial order, $n=1$ (see section S5, SI for further details on higher order WGMs).

In all cases across the parametric sweep, we evaluate the system against the conventional strong coupling criterion, $2g > (\gamma_{exc} + \kappa_j)$ \cite{Vahala2003}. For the representative $\text{TE}_{18,1}$ mode of the 2.5~$\mu$m microsphere, $2g = 121.98$~meV substantially exceeds $(\gamma_{exc} + \kappa_{18,1}) = 73.54$~meV, and an equivalent inequality holds for every interacting mode across the entire 1.5-4.0~$\mu$m sweep. By this conventional benchmark, the system is firmly within the strong coupling regime at every diameter examined.

This uniform satisfaction of the threshold, however, masks a non-trivial morphological scaling that this criterion itself is insensitive to. Even though the coupling strength $g$ reduces to half and $\kappa$ varies nearly an order of magnitude across the swept diameters, the conventional inequality being an average of the two loss channels, obscures the underlying competition between coherent exchange and dissipation. Physically, a polariton corresponds to energy oscillating back and forth between the photonic and excitonic components of the hybrid state at a rate governed by $g$, with the cavity leaking photons at rate $\kappa$ and the molecular ensemble dephasing at rate $\gamma_{exc}$. The coherent exchange is therefore limited by the lossier channel. The faster channel wins the race against $g$ and sets the polariton lifetime. The physically relevant comparison is thus not between $2g$ and the sum of $\kappa$ and $\gamma_{exc}$, but between $g$ and the larger of the two losses. The mean loss understates this bottleneck whenever the two channels are asymmetric, as is generically the case for organic emitters coupled to dielectric cavities. A cavity in which $g$ outpaces this dominant dissipation channel sustains a greater number of coherent exchanges before the hybrid state decays, yielding longer polariton lifetimes and hybrid eigenstates with more pronounced mixed light-matter character. To identify the cavity geometry that best satisfies this stricter condition, the continuous evolution of $g$, $\kappa$, and $\gamma_{exc}$ must be examined together across the full morphological sweep.

\begin{figure}[htbp]
  \centering
  \includegraphics[width=1.0\textwidth]{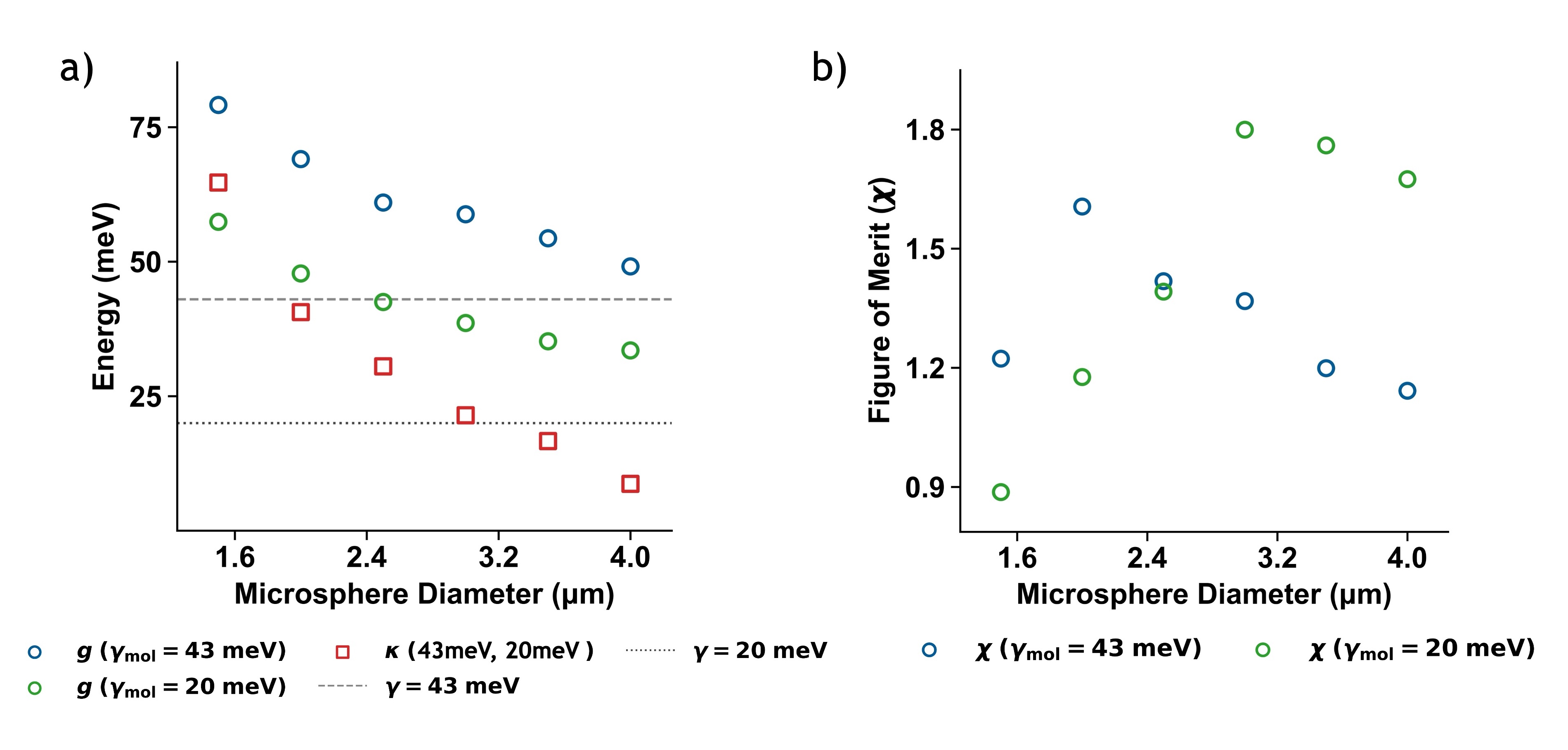}
  \caption{Evolution of the extracted coupling and loss parameters and the resulting Hybrid Figure of Merit ($\chi$) as a function of microsphere diameter. (a) Coupling strength $g$ (blue and green circles) and cavity linewidth $\kappa$ (red squares) extracted from the COM fits, plotted alongside the intrinsic molecular linewidth $\gamma_{exc}$ (dashed and dotted horizontal lines for 43~meV and 20~meV respectively). (b) The Hybrid Figure of Merit $\chi = g/\mathrm{max}(\kappa, \gamma_{exc})$ for both molecular scenarios ($\gamma_{exc}$ = 43~meV and 20~meV), reaching maximum when the diameter where the cavity linewidth matches the molecular linewidth.}
  \label{fig:fom}
\end{figure}

Figure~\ref{fig:fom}a compiles the COM-extracted parameters across the full sweep. Two competing trends are evidently seen. The coupling strength $g$ decreases with diameter, from $\sim$80~meV at 1.5~$\mu$m to $\sim$49.12~meV at 4.0~$\mu$m, a scaling that follows directly from the $g \propto 1/\sqrt{V}$ dependence\cite{Vasista2020}. As the microsphere expands, the WGM mode volume grows as $\sim R^2$, where $R$ is the radius of the sphere making the coupling strength, $g$, scale as $\sim \frac{1}{R}$. The molecular linewidth $\gamma_{exc} = 43$~meV (dashed line) is a material constant, it is set by the dephasing within the TDBC J-aggregate and is wholly independent of cavity geometry.

Once $\kappa$ falls below $\gamma_{exc}$ (here, beyond $\sim$2.0~$\mu$m), the cavity is no longer the limiting dissipation channel. The polariton coherence is bottlenecked by molecular dephasing, and further reductions in $\kappa$ produce no corresponding gain in hybrid-state quality. Yet each incremental increase in $R$ continues to penalize $g$ through mode volume dilution. Optimization in this regime is thus counterproductive. The cavity-centric FOM rewards a parameter ($\kappa$) whose continued minimization yields no physical benefit after a certain point, while ignoring the simultaneous suppression of the coupling strength that actually governs the energy exchange rate.

This asymmetry calls for a figure of merit that evaluates the coherent coupling against the system's dominant decoherence channel rather than against the sum of dissipation rates. Hence we define
\begin{equation}
\chi = \frac{g}{\mathrm{max}(\kappa, \gamma_{exc})},
\end{equation}
a dimensionless quantity that explicitly identifies which loss process limits the polariton lifetime for a given geometry. The structure of $\chi$ has a direct physical interpretation, at small diameters where $\kappa > \gamma_{exc}$, cavity leakage dominates and $\chi$ improves as the resonator quality factor increases. At large diameters where $\kappa < \gamma_{exc}$, molecular dephasing dominates, the denominator collapses to the constant $\gamma_{exc}$, and $\chi$ tracks $g$ directly, so any further increase in $R$ degrades the metric in lockstep with the loss in coupling strength. The crossover, $\kappa \approx \gamma_{exc}$, is therefore the unique geometry at which neither dissipation channel is being needlessly over-engineered and the coherent exchange rate is maximized against the bottleneck that actually limits it.

The behavior of $\chi$ across our swept geometries is plotted in Fig.~\ref{fig:fom}b. For the TDBC case ($\gamma_{exc} = 43$~meV), $\chi$ peaks at diameter $D \approx 2.0~\mu$m, precisely where the extracted cavity linewidth $\kappa \approx 40.6$~meV most closely matches the molecular linewidth. At this morphology, $g \approx 69$~meV is retained at $\sim$85\% of its peak value (observed at $D = 1.5~\mu$m) while $\kappa$ has been reduced to the molecular floor; pushing to larger diameters drops $\kappa$ further to no useful effect, and $\chi$ falls thereafter as $g$ continues to dilute. To confirm that this peak reflects the matching condition $\kappa(D) \approx \gamma_{exc}$ rather than a coincidence of the TDBC parameters, we repeated the full parametric sweep with the molecular linewidth in the Lorentz model artificially reduced to $\gamma_{exc} = 20$~meV, keeping the value of permittivity value the same as that for $\gamma_{exc}=43$~meV. As $g \propto \sqrt{f}$, $f$ being the oscillator strength, the coupling strength $g$ reduces as seen in figure \ref{fig:fom} (a). Additionally, the $\chi$ maximum shifts, as predicted, to $D \approx 3.0~\mu$m, the diameter at which the cavity linewidth crosses the new molecular floor (Fig.~\ref{fig:fom}b, blue open circles). The matching condition $\kappa \approx \gamma_{exc}$ is thus the operative design principle, independent of the specific emitter.
\begin{comment}
    This result inverts the standard cavity design heuristic for morphology-dependent resonators coupled to dissipative molecular ensembles. The conventional $Q/\sqrt{V}$ optimization is a cavity-only metric. It promotes conditions under which driving $\kappa \to 0$ prolongs the polariton lifetime. The TDBC linewidth $\gamma_{exc}$ is fixed by intramolecular vibrational coupling and cannot be reduced by cavity engineering, and the WGM geometry couples $\kappa$ and $V$ through the single parameter $R$, so any reduction in $\kappa$ is necessarily accompanied by a dilution of $g$. The $\chi$ metric resolves this by promoting the emitter linewidth from an external constant to an active term in the optimization picture. The cavity is no longer designed in isolation and then made to interact with a molecule, but selected such that its photonic loss rate matches the dephasing rate of the molecule it will host.
\end{comment}

Operationally, this collapses cavity selection to a single, fabrication-free criterion. Given a target molecule with known $\gamma_{exc}$, the optimal diameter $D$ is when $\kappa(D) = \gamma_{exc}$, where $\kappa(D)$ is the bare-cavity linewidth obtained either from a single FEM sweep of uncoated microspheres or directly from dark-field scattering measurements on commercially available samples. This protocol is particularly relevant for applications such as polaritonic chemistry and cavity-enhanced sensing, in which the emitter is dictated by the chemistry of interest and the cavity must be tuned to it, rather than vice versa. For the sake of completeness, we also calculated cooperativity, $C$, for the TDBC coupled soft-cavities and found that $C$ monotonically increases as we increase the diameter of the cavity (See section S4, SI). This monotonic increase is the direct impact of the sharp reduction in $\kappa$ as one increases the diameter of the microsphere.

In summary, we systematically studied TDBC coated soft-cavities with an aim to achieve robust and efficient strong coupling. We introduced a modified figure of merit parameter ($\chi$) which encapsulates the effect of competing loss mechanisms, particularly their asymmetricity, on the molecule-cavity coupling. We show that the $\chi$ is maximum when the molecule and cavity linewidths are equal. The condition $\kappa \approx \gamma$ can be interpreted as a dissipation-matching condition analogous to impedance matching in coupled oscillatory systems. When cavity and molecular losses become comparable, neither subsystem acts as a dominant sink for coherence, thereby maximizing the robustness of reversible light--matter exchange. We believe that this study opens up a new avenue in understanding and designing chemically open optical cavities for molecular strong coupling. A deeper mechanistic understanding of the dissipation-matching condition and its connection to light--matter dynamics in multimodal cavities warrants further investigation, but lies beyond the scope of the present work.

\section*{Acknowledgements}
This research was partially funded by Prime Minister's Early Career Research Grant (PMECRG) from Anusandhan National Research Foundation (grant no. ANRF/ECRG/2024/006239/PMS). The authors acknowledge the support provided by the Indian Science Technology and Engineering facilities Map (I-STEM), an initiative of the Office of the Principal Scientific Adviser, Government of India, for granting access to COMSOL Multiphysics software suite. ABV acknowledges the support from IISER Bhopal research initiation grant.

\section*{Author Contributions}
SMG performed all the numerical simulations and analyses and co-wrote the manuscript. ABV conceptualized the idea, supervised the project and co-wrote the manuscript.

%%%%%%%%%%%%%%%%%%%%%%%%%%%%%%%%%%%%%%%%%%%%%%%%%%%%%%%%%%%%%%%%%%%%%
%% References
%%%%%%%%%%%%%%%%%%%%%%%%%%%%%%%%%%%%%%%%%%%%%%%%%%%%%%%%%%%%%%%%%%%%%
\bibliographystyle{unsrtnat}
\bibliography{References}

%%%%%%%%%%%%%%%%%%%%%%%%%%%%%%%%%%%%%%%%%%%%%%%%%%%%%%%%%%%%%%%%%%%%%
%% Supplementary Information
%% Begins after the references. Figures and sections are S-prefixed
%% and counters are reset so that the SI is self-contained.
%%%%%%%%%%%%%%%%%%%%%%%%%%%%%%%%%%%%%%%%%%%%%%%%%%%%%%%%%%%%%%%%%%%%%
\clearpage

% Re-enable section numbering for the SI and switch to the S-prefixed scheme.
\setcounter{secnumdepth}{3}
\setcounter{section}{0}
\setcounter{figure}{0}
\setcounter{equation}{0}
\renewcommand{\thesection}{S\arabic{section}}
\renewcommand{\thesubsection}{S\arabic{section}.\arabic{subsection}}
\renewcommand{\thefigure}{S\arabic{figure}}
\renewcommand{\theequation}{S\arabic{equation}}

\begin{center}
  {\large\bfseries Supplementary Information\par}
  \vspace{0.5em}
  {\large Tailoring soft cavities for robust molecular strong coupling\par}
\end{center}
\vspace{1em}

\section{Two-Dimensional FEM Modeling of the Microsphere System}

Two-dimensional finite element method (FEM) simulations were performed in COMSOL Multiphysics (Wave Optics module, Electromagnetic Waves, Frequency Domain interface) to compute the optical response of TDBC-coated polystyrene microspheres positioned on a glass substrate. The 2D cross-sectional geometry, rather than a full 3D model, was chosen to capture the relevant whispering gallery mode (WGM) physics at substantially reduced computational cost. The evanescent coupling and near-field light-matter exchange that govern strong coupling in this system are accurately reproduced in the equatorial cross-section.

\subsection{Geometry and material definitions}
The simulation domain consists of an upper air half-space and a lower glass half-space separated by a horizontal interface. A circular polystyrene microsphere of variable diameter is positioned on the glass-air boundary. A concentric shell of fixed thickness 8~nm, representing four molecular layers of TDBC J-aggregate deposited via layer-by-layer self-assembly, is added to the microsphere exterior. This shell thickness is held constant across the entire parametric sweep, consistent with the deposition method characterized in prior experimental work \cite{Vasista2020}. Polystyrene, glass, and air are assigned constant real refractive indices of 1.59, 1.52, and 1.0 respectively across the simulated wavelength range. The TDBC shell is modeled as a dispersive Lorentz oscillator with the parameters given in Section S2. This captures the macroscopic excitonic absorption of the J-aggregate at $\lambda_{\mathrm{exc}} = 582$~nm with intrinsic linewidth $\gamma_{\mathrm{exc}} = 43$~meV.

\subsection{Excitation: evanescent coupling via frustrated TIR}
The microsphere is excited in the evanescent configuration that mirrors the experimental Dove-prism geometry. A transverse-electric (TE) polarized broadband plane wave is launched from the bottom-left of the glass domain via an active scattering boundary condition (SBC), at an incident angle of $\theta_i = 47^\circ$, exceeding the critical angle for the glass-air interface ($\theta_c \approx 41.1^\circ$). The resulting total internal reflection at the glass-air boundary generates an exponentially decaying evanescent field on the air side, into which the microsphere is placed. We restrict the analysis to TE modes by selecting the polarization of the incident field. All remaining external edges of the rectangular simulation domain are assigned passive scattering boundary conditions to absorb outgoing radiation and suppress unphysical back-reflections.

\begin{figure}[htbp]
  \centering
  \includegraphics[width=0.95\textwidth]{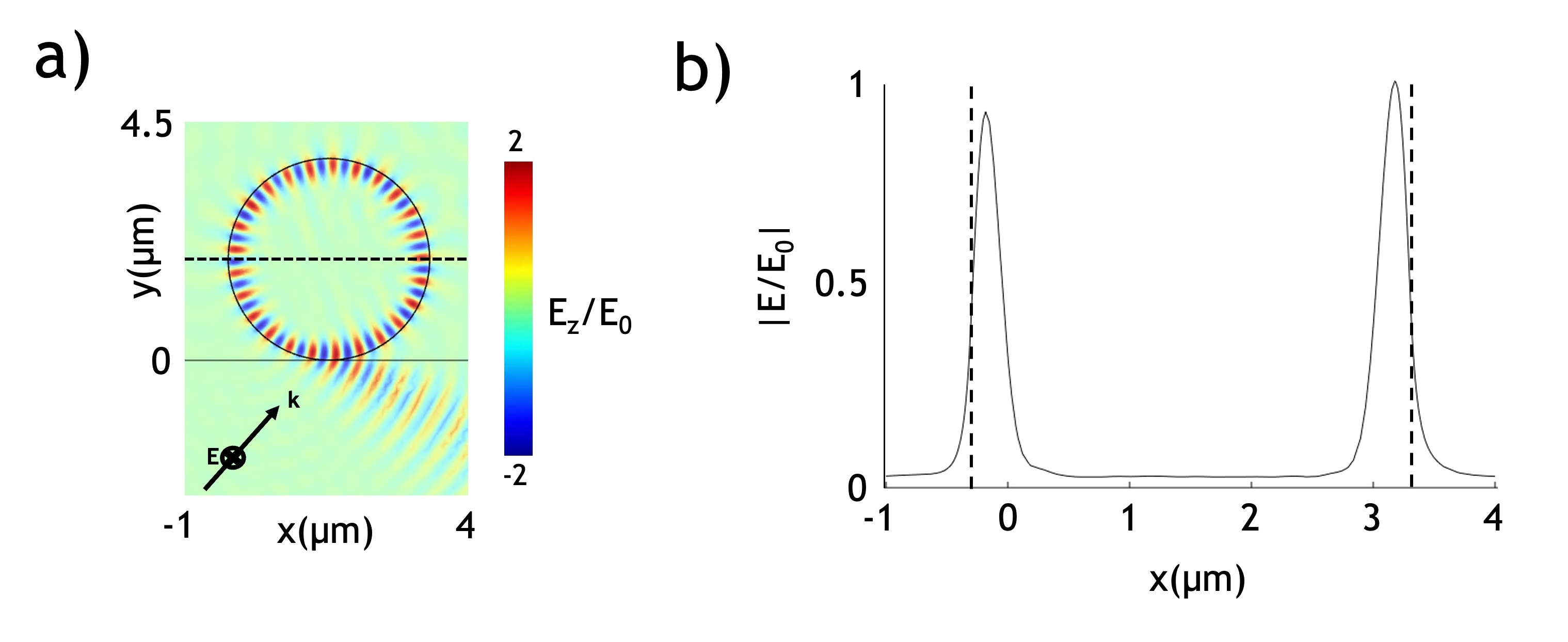}
  \caption{Simulated whispering gallery mode in a bare 3.5~$\mu$m polystyrene microsphere on a glass substrate. (a) Out-of-plane electric field $E_z/E_0$ at a representative TE resonance. The incident TE-polarized plane wave is launched from below at $\theta_i = 47^\circ$ (above the critical angle), generating the evanescent field that drives the WGM via frustrated total internal reflection. This indicates the microsphere boundary. (b)Radial line profile of the normalized field magnitude $|E/E_0|$ along the equatorial cut indicated by the dotted line in (a). The two sharp peaks at the microsphere boundary, with negligible field amplitude in the bulk interior, confirm the fundamental radial order ($n = 1$) of the excited mode and the strong field localization at the outer dielectric surface where the TDBC shell resides.}
  \label{fig:wgm_field}
\end{figure}

Figure~\ref{fig:wgm_field} shows the simulated $E_z/E_0$ distribution at a representative WGM resonance of a 3.5~$\mu$m bare microsphere, illustrating the characteristic azimuthal field pattern at the dielectric periphery and the excitation. The radial line profile of $|E/E_0|$ extracted along the equatorial axis (Fig.~\ref{fig:wgm_field}b) confirms the extreme spatial confinement of the fundamental radial mode ($l = 1$) at the outer dielectric boundary, where the molecular shell is located.

\subsection{Meshing strategy}
The geometry presents a multi-scale meshing challenge: an 8~nm molecular shell concentric with a micron-scale dielectric sphere, with a sub-nanometric tangent contact between sphere and substrate. A spatially graded meshing strategy was therefore used. Within the molecular shell, along the full circumference of the microsphere, and in the immediate vicinity of the sphere-glass contact, a fine triangular mesh with element size below 5~nm was enforced to resolve the steep field gradients in the evanescent tail. The mesh density was relaxed progressively into the background air and glass domains, where the field is comparatively delocalized, reaching element sizes on the order of $\lambda/6$ at the outermost boundaries. This gradient approach preserved physical accuracy in the optically active regions while keeping the total degree-of-freedom count tractable across the full parametric sweep.

\subsection{Data extraction: internal field integral as scattering proxy}
For each simulated geometry, the optical response was quantified by the two-dimensional surface integral of the electric field magnitude $|E|^2$ over the microsphere interior.
At resonance, near-field coupling drives a substantial buildup of stored electromagnetic energy inside the cavity. Since radiative emission scales linearly with the stored energy, the internal field integral serves as a proxy for the measured far-field scattering cross-section, without requiring an explicit far-field projection. This choice considerably simplifies the post-processing pipeline while preserving the spectral information needed to identify WGM resonance positions and linewidths.

A background-subtraction step was applied to suppress the contribution of light scattered directly off the bare glass-air interface, which would otherwise reduce the signal-to-noise ratio of the WGM peaks. Each simulation was run twice, once with the microsphere domain disabled (yielding the background field of the bare TIR interface) and once with the full microsphere geometry in place. The background was subtracted from the full-geometry result before the surface integral was evaluated.

\subsection{Mode indexing}
Whispering gallery modes are classified by their electric-field polarization as transverse-electric (TE) or transverse-magnetic (TM), and are further labeled by three indices $(n,\, l,\, m)$ \cite{Vasista2020}. The radial number $n$ counts the intensity maxima along the radial direction, the orbital number $l$ is half the number of intensity maxima around the circumference, and the azimuthal number $m \in \{-l,\dots,+l\}$ gives the projection of $l$ onto the quantization axis. In the 2D cross-section of Fig.~\ref{fig:wgm_field}, the radial profile (panel b) shows a single field maximum on each side of the sphere boundary, identifying the excited mode as the fundamental radial order $n = 1$; the azimuthal pattern (panel a) exhibits $2l=50$ intensity lobes around the circumference, corresponding to $l=25$ for this 3.5~$\mu$m TE resonance.

\section{Lorentzian Permittivity Model of the TDBC Molecular Layer}

The TDBC J-aggregate shell coating the polystyrene microsphere was modeled as a single Lorentzian oscillator with complex relative permittivity
\begin{equation}
\varepsilon_{\mathrm{TDBC}}(E) = \varepsilon_{\infty} + \frac{f E_{\mathrm{res}}^{2}}{E_{\mathrm{res}}^{2} - E^{2} - i E \gamma_{\mathrm{res}}},
\label{eq:lorentz}
\end{equation}
where $\varepsilon_{\infty}$ is the background permittivity, $E_{\mathrm{res}}$ is the excitonic resonance energy, $\gamma_{\mathrm{res}}$ is the resonance linewidth, and $f$ is the reduced oscillator strength. The thickness of a single molecular layer of the TDBC system is taken to be $\sim$2~nm, as established in prior work on layer-by-layer self-assembled films \cite{vasista_2020_SI_ref1}; the four-bilayer shell used throughout this study therefore corresponds to approximately a total thickness of 8~nm.

\begin{figure}[htbp]
  \centering
  \includegraphics[width=0.75\textwidth]{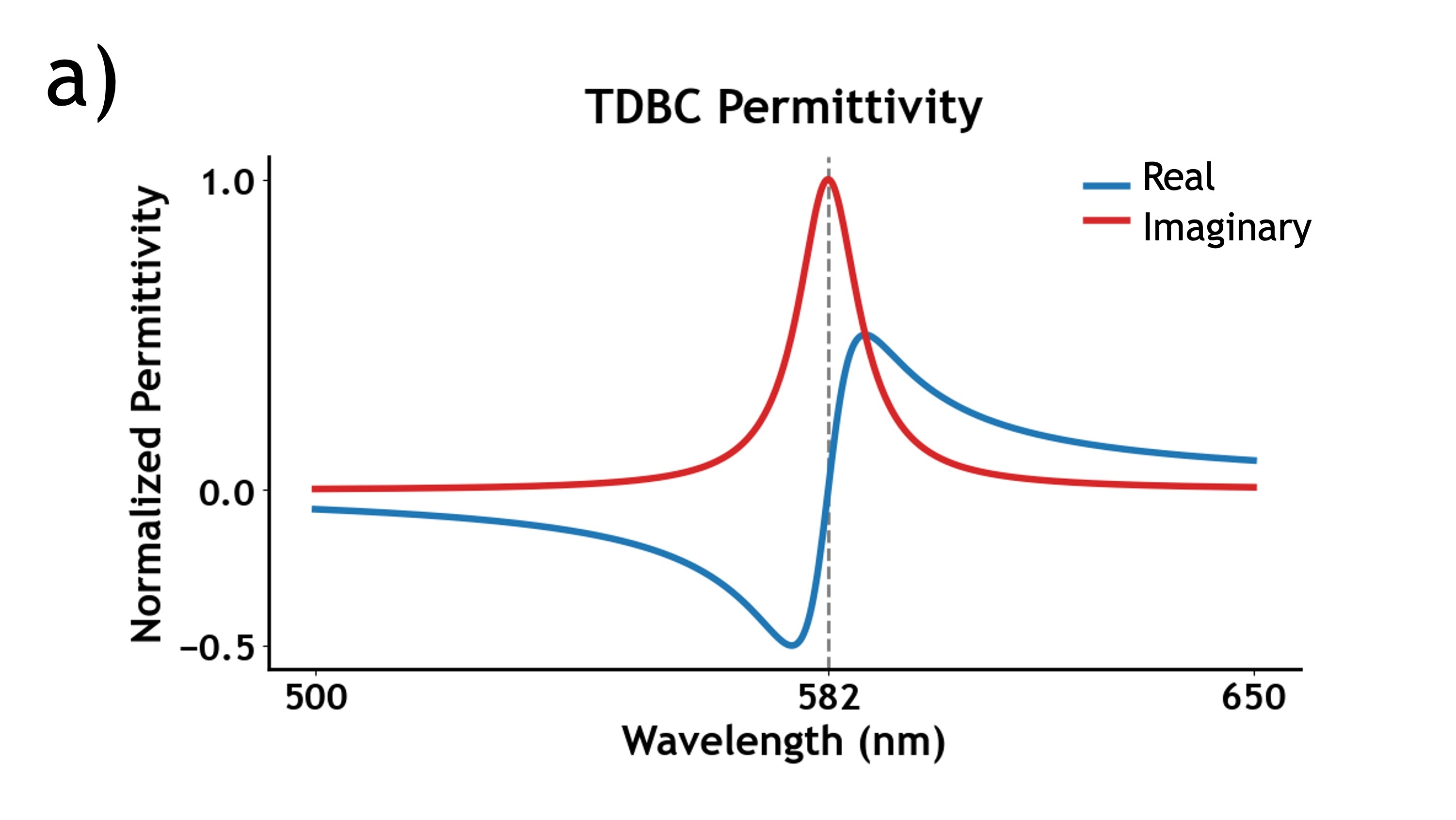}
  \caption{Normalized real (blue) and imaginary (red) components of the complex relative permittivity $\varepsilon_{\mathrm{TDBC}}(E)$ of the TDBC J-aggregate shell, computed from Eq.~\eqref{eq:lorentz} with the parameters listed in the text. The dashed vertical line marks the excitonic resonance at $\lambda_{\mathrm{res}} = 582$~nm. The peak in $\mathrm{Im}[\varepsilon]$ at resonance corresponds to the J-band absorption maximum; the linewidth of this peak is $\gamma_{\mathrm{res}} = 43$~meV, which sets the molecular dissipation rate $\gamma_{\mathrm{exc}}$ entering the coupled oscillator model.}
  \label{fig:tdbc_permittivity}
\end{figure}

The parameters used in Eq.~\eqref{eq:lorentz} are $\varepsilon_{\infty} = 2.36$, $f = 0.36$, and $\gamma_{\mathrm{res}} = 43$~meV, corresponding to the intrinsic dephasing rate of the TDBC J-band at room temperature. The resonance energy was set to $E_{\mathrm{res}} = 2.13$~eV ($\lambda_{\mathrm{res}} = 582$~nm), which matches the absorption maximum of PDAC/TDBC films deposited on dielectric surfaces and accounts for the slight substrate-induced shift relative to monolayer TDBC on quartz \cite{Vasista2020}. The resulting real and imaginary components of the complex permittivity are plotted in Fig.~\ref{fig:tdbc_permittivity}, showing the characteristic dispersive (real) and absorptive (imaginary) Lorentzian profiles centered at the excitonic resonance. For the case of $\gamma_{res}=20$~meV, we reduce the oscillator strength, $f$, to 0.167 so as to make the overall permittivity values equal to the $\gamma_{res}=43$~meV case.

\section{Hopfield Coefficients of the Polaritonic States}

Diagonalization of the Hermitian part of the coupled oscillator Hamiltonian (Eq.~(2) of the main text) yields, in addition to the polariton eigenenergies, the eigenvectors describing the photonic and excitonic content of each hybrid state. For a given polariton branch, the squared eigenvector components $|\alpha|^2$ and $|\beta|^2$ quantify the cavity and molecular fractions respectively, and satisfy $|\alpha|^2 + |\beta|^2 = 1$ at every detuning. At zero detuning, where the bare cavity mode is degenerate with the excitonic resonance, both polariton branches are equally photonic and excitonic ($|\alpha|^2 = |\beta|^2 = 0.5$); away from resonance, the upper and lower polaritons asymptotically inherit the character of the bare state to which they are closer in energy.

\begin{figure}[htbp]
  \centering
  \includegraphics[width=\textwidth,height=0.85\textheight,keepaspectratio]{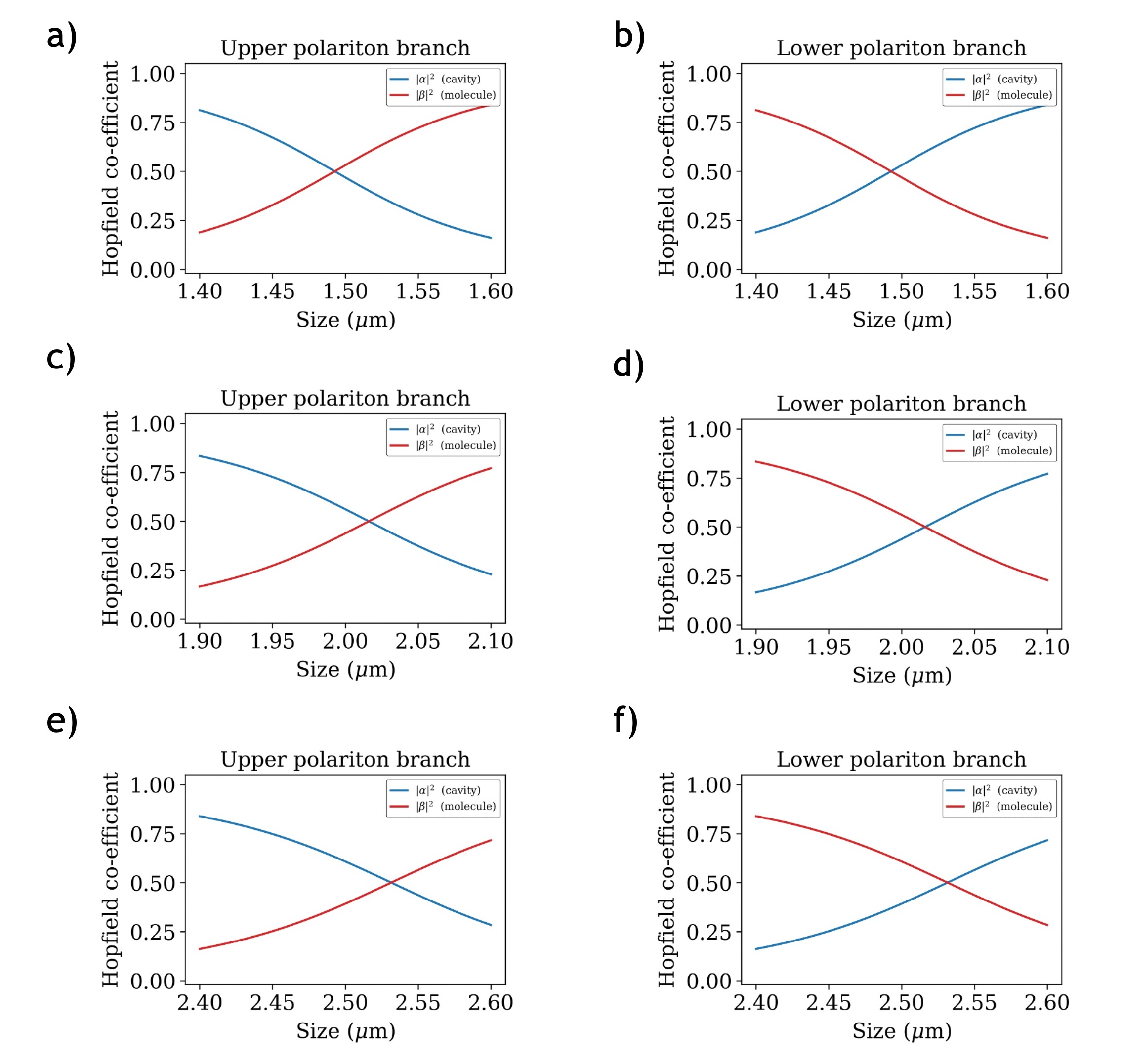}
  \caption{Hopfield coefficients $|\alpha|^2$ (cavity, blue) and $|\beta|^2$ (molecule, red) for the upper and lower polariton branches as a function of microsphere diameter, extracted from the diagonalization of the coupled oscillator Hamiltonian using the coupling strengths $g$ obtained from the main-text fits. (a, b) Upper and lower polariton branches for the 1.5~$\mu$m microsphere, mode $TE_{10,1}$. (c, d) Same for the 2.0~$\mu$m microsphere, mode $TE_{14,1}$. (e, f) Same for the 2.5~$\mu$m microsphere, mode $TE_{18,1}$. In every panel, the two coefficients cross at $|\alpha|^2 = |\beta|^2 = 0.5$ at the resonance diameter where $E_{\mathrm{cav}}(D) = E_{\mathrm{exc}}$, confirming the expected 50/50 photon-exciton mixing at zero detuning.}
  \label{fig:hopfield}
\end{figure}

Because the WGM resonators studied here are morphology-dependent, the cavity-exciton detuning is tuned not by an external wavevector but by the microsphere diameter, which shifts $E_{\mathrm{cav}}(D)$ across the stationary excitonic energy $E_{\mathrm{exc}}$. The Hopfield coefficients consequently trace out the cavity-molecule mixing as a continuous function of $D$. Figure~\ref{fig:hopfield} presents the computed coefficients for the three microsphere diameters that bracket the optimum identified by the hybrid figure of merit $\chi$ in the main text. For each diameter, both interacting TE modes (Pair 1 and Pair 2, as defined in the main-text fits) are shown, with the upper and lower polariton branches plotted separately.

\section{Cooperativity and its Comparison with the Hybrid Figure of Merit $\chi$}

The standard figure of merit for the dissipative side of cavity quantum electrodynamics is the single-emitter cooperativity,
\begin{equation}
C = \frac{g^{2}}{\kappa \, \gamma_{\mathrm{exc}}},
\label{eq:cooperativity}
\end{equation}
where $\kappa$ and $\gamma_{\mathrm{exc}}$ are the cavity and molecular FWHM linewidths \cite{Torma2015,Vahala2003}. Physically, $C$ counts the number of coherent absorption-and-re-emission cycles a single excitation completes inside the cavity before being lost to either dissipation channel, and $C > 1$ is the conventional threshold for the onset of strong coupling.

\begin{figure}[htbp]
  \centering
  \includegraphics[width=0.75\textwidth]{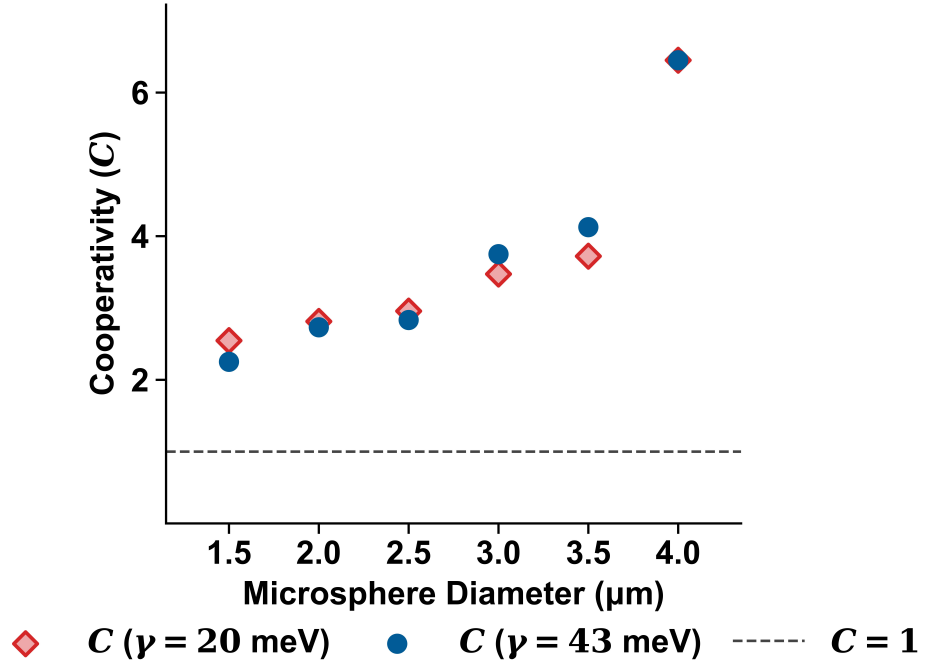}
  \caption{Cooperativity $C = g^{2}/(\kappa \gamma_{\mathrm{exc}})$ as a function of microsphere diameter for both molecular scenarios studied in the main text: $\gamma_{\mathrm{exc}} = 43$~meV (blue circles) and $\gamma_{\mathrm{exc}} = 20$~meV (red squares). The dashed horizontal line marks the conventional strong-coupling threshold $C = 1$. In contrast to the hybrid figure of merit $\chi$ presented in the main text, $C$ rises monotonically with diameter and exhibits no extremum, since the product $\kappa \gamma_{\mathrm{exc}}$ continues to decrease even after the cavity linewidth has fallen below the molecular floor.}
  \label{fig:cooperativity}
\end{figure}

Cooperativity weights $\kappa$ and $\gamma_{\mathrm{exc}}$ symmetrically through their product, so it does not distinguish whether the polariton lifetime is bottlenecked by cavity leakage or by molecular dephasing. Once the smaller of the two linewidths falls below the larger, further reductions of it cease to improve the polariton coherence, yet $C$ continues to grow because the denominator $\kappa \gamma_{\mathrm{exc}}$ continues to shrink. For the soft-cavity geometries studied here, where $\gamma_{\mathrm{exc}}$ is fixed by intramolecular vibrational coupling and $\kappa(R)$ is the single tunable knob, this means that $C$ rises monotonically with diameter (Fig.~\ref{fig:cooperativity}) and never indicates an optimal cavity size, even though the coupling strength $g$ continues to dilute through mode-volume scaling.

The hybrid figure of merit $\chi = g / \max(\kappa, \gamma_{\mathrm{exc}})$ used in the main text is constructed to remove this blindness. Replacing the symmetric product in $C$ with the asymmetric $\max(\kappa, \gamma_{\mathrm{exc}})$ encodes the premise that the polariton coherence is rate-limited by whichever loss channel is faster, and that reducing the slower channel below the faster one yields no further coherence. As a result, the two metrics answer different questions: cooperativity benchmarks the global strength of a coupled system against the strong-coupling threshold; $\chi$ identifies the optimal cavity within a single, continuously tuned architecture by promoting the dissipation-matching condition $\kappa \approx \gamma_{\mathrm{exc}}$ to an explicit feature of the metric. For the design question relevant to morphology-dependent soft cavities, where the cavity must be tuned to a chemically-fixed emitter, $\chi$ is therefore the more informative design parameter, while $C$ remains the correct figure of merit for cross-platform comparison.

\section{Higher-Order Mode Coupling}

To understand higher order mode coupling without any information loss, the dissipation of the coated molecule in the COMSOL simulations was artificially reduced to 10 meV for a larger $4\mu m$ microsphere. For these microspheres, a second whispering gallery resonance with a different radial order $n$ falls within the spectral window of the TDBC excitonic transition and couples to the molecular shell in parallel with the fundamental $n = 1$ mode (See Figure S5 (c). The fundamental $TE_{30,1}$ mode (Pair 2, blue) (Figure S5 (b)) has its field maximum at the outer dielectric surface where the molecular shell resides, while the higher-order radial $TE_{26,2}$ mode (Pair 1, green) (Figure S5 (a)) has additional intensity lobes pushed inward into the polystyrene bulk and a comparatively reduced amplitude at the shell. At the same time, the fundamental-order mode has a longer photon lifetime inside the sphere, giving a narrower cavity linewidth $\kappa$. For the 4~$\mu$m microsphere, this trade-off slightly favors the fundamental-order mode: the higher-order mode (Pair 1) yields $g = 23.25$~meV at $\kappa = 8.70$~meV, while the fundamental-order mode (Pair 2) yields a comparable $g = 24.27$~meV at $\kappa = 6.17$~meV. The competition between the shell-overlap and reduced cavity dissipation thus produces two nearly equivalent strong-coupling channels in the same resonator, the fundamental-order mode being marginally more robust due to its narrower linewidth.

\begin{figure}[htbp]
  \centering
  \includegraphics[width=0.75\textwidth]{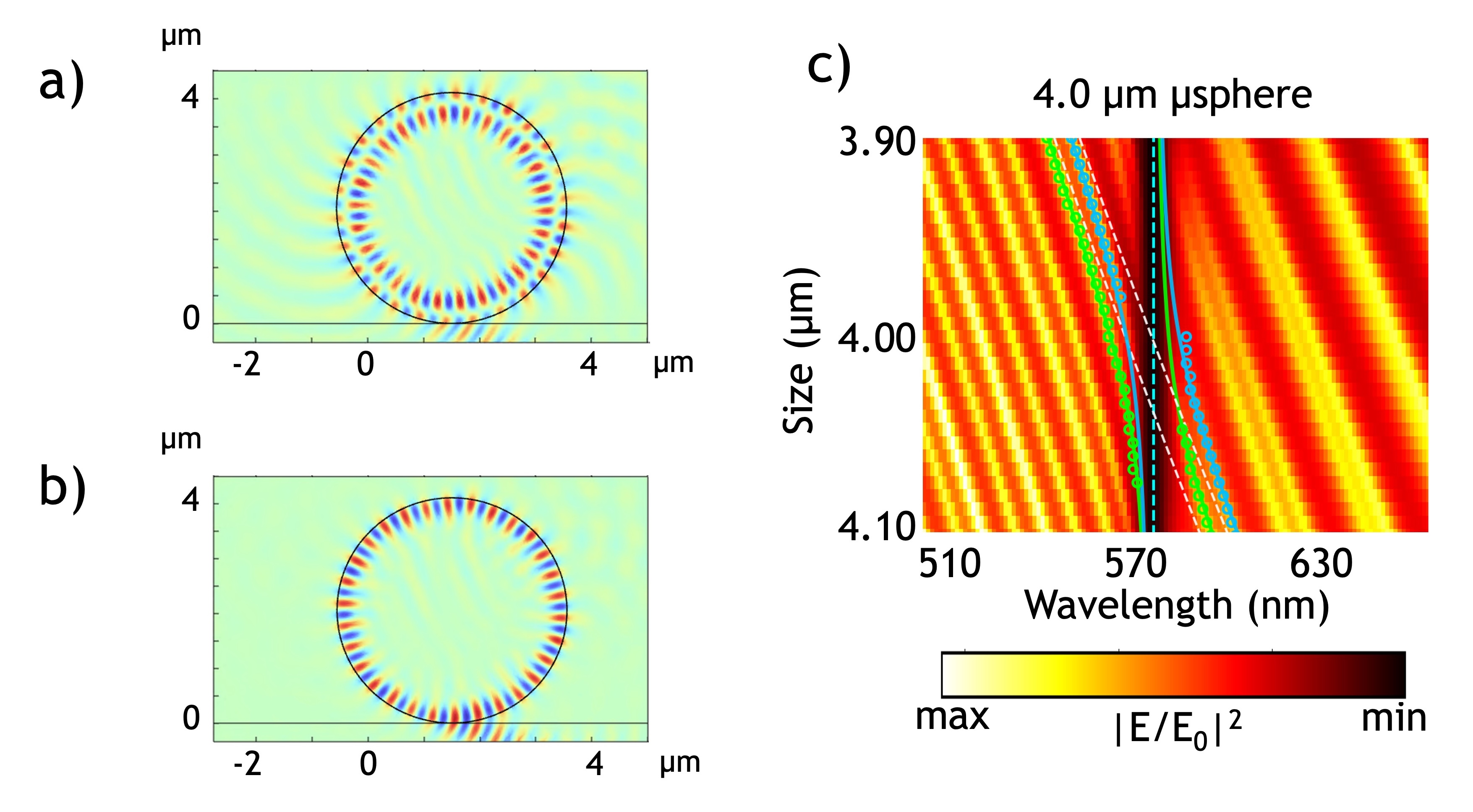}
  \caption{Simultaneous fit of the two co-existing TE polariton pairs in a 4~$\mu$m microsphere. Pair 1 (green, radial order $n = 2$, $TE_{26,2}$) and Pair 2 (blue, fundamental order radial mode $n=1$, $TE_{30,1}$) both undergo strong coupling to the TDBC excitonic transition. The extracted parameters are $g = 23.25$~meV, $\kappa = 8.70$~meV for Pair 1 and $g = 24.27$~meV, $\kappa = 6.17$~meV for Pair 2.}
  \label{fig:ho-modes}
\end{figure}

\end{document}